\documentclass[journal,10pt]{IEEEtran}
\pdfoutput=1
\usepackage{cite}

\usepackage{amsmath}
\usepackage{algorithm}
\usepackage{algorithmic}
\usepackage{subfigure}
\usepackage{amssymb}
\usepackage{graphicx}
\usepackage{empheq}

\usepackage{array}
\usepackage{mathtools}
\usepackage{graphicx}
\usepackage{cases}
\usepackage[colorinlistoftodos]{todonotes}
\usepackage[colorlinks=true, linkcolor=black,citecolor=black,urlcolor=black]{hyperref}
\usepackage{indentfirst}
\usepackage{multirow}
\usepackage{booktabs}
\usepackage{makecell}
\allowdisplaybreaks[4]

\usepackage[switch]{lineno}
\usepackage{changes}

\usepackage{url}
\usepackage[hyphenbreaks]{breakurl}

\begin{document}
\title{Robust Deployment and Resource Allocation for Robotic Aerial Base Station Enabled OFDM Integrated Sensing and Communication \\
}


\author{
Yuan~Liao, \IEEEmembership{Student Member,~IEEE}, Vasilis~Friderikos, \IEEEmembership{Member,~IEEE}, Halim Yanikomeroglu,  \IEEEmembership{Fellow,~IEEE}

\thanks{Yuan Liao and Vasilis Friderikos are with the Department of Engineering, King's College London, London WC2R 2LS, U.K. (e-mail: yuan.liao@kcl.ac.uk; vasilis.friderikos@kcl.ac.uk).

Halim Yanikomeroglu is with the Non-Terrestrial 
Networks (NTN) Lab, Department of Systems and Computer Engineering, Carleton University, Ottawa, ON K1S 5B6, Canada (e-mail: halim@sce.carleton.ca).}
}


\maketitle

\begin{abstract}
The envisioned robotic aerial base station (RABS) concept is expected to bring further flexibility to integrated sensing and communication (ISAC) systems. In this letter, characterizing the spatial traffic distribution on a grid-based model, the RABS-assisted ISAC system is formulated as a robust optimization problem to maximize the minimum satisfaction rate (SR) under a cardinality constrained uncertainty set. The problem is reformulated as a mixed-integer linear programming (MILP) and solved approximately by the iterative linear programming rounding algorithm. Numerical investigations show that the minimum SR can be improved by 28.61\% on average compared to fixed small cells.

\end{abstract}


\vspace{-0.2cm}
\section{Introduction}

In the upcoming 6G era, reliable wireless coverage and accurate remote sensing capability are crucial for emerging applications such as intelligent transport systems and smart manufacturing. This has led to the recent surge in the development of integrated sensing and communication (ISAC) techniques. To enhance the flexibility and adjustability of ISAC systems, in this paper, we employ robotic aerial base stations (RABS) that can attach autonomously to lampposts or other tall urban landforms via energy neutral grasping, and fly to another grasping point via controllable maneuverability to perform the sensing and communication functions. 

A number of works are devoted to perform ISAC tasks to improve spectrum efficiency and reduce the expenditure cost. In \cite{liu2017adaptive}, the sensing and communication performances, evaluated by mutual information (MI) and data rate respectively, are maximized jointly under the limitation of transmission power. The work \cite{liu2019robust} extends this approach by incorporating channel uncertainty, while in \cite{shi2018low}, the transmission power is minimized while ensuring predefined thresholds for both MI and data rate. The subcarrier assignment problem is considered in \cite{shi2019joint,nguyen2021hierarchical} to optimize the transmission power and satisfaction utility, respectively. Besides the conventional terrestrial cells, unmanned aerial vehicle (UAV) is expected to improve the flexibility of next generation cellular networks \cite{demirtas2022deep}. The work \cite{cui2023toward} employs UAVs to perform ISAC tasks to improve the security and reliability of networks. The communication throughput and energy efficiency are optimized in the UAV-assisted ISAC systems in \cite{meng2022uav,qin2023deep}, respectively. However, to overcome the issue that the serving endurance of UAVs is severely confined by the on-board battery capacity, the work \cite{liao2023optimal} proposes the prototype of RABS carried by an UAV and mounted with a mechanical grasper so that it can attach on lampposts when providing wireless coverage and agilely relocate to another hot-spot to adapt to the traffic dynamic. The service time is significantly increased due to the lower grasping power (tens of Watts) compared to the hovering/flying power of UAV base stations (hundreds of Watts) \cite{liao2023optimal}.

In this letter, we employ the RABS to perform ISAC tasks in a flexible and energy-efficient manner. Moreover, instead of assuming that the users' locations are fixed and known as \cite{liu2017adaptive,liu2019robust,shi2018low,shi2019joint,nguyen2021hierarchical,demirtas2022deep,cui2023toward,meng2022uav,qin2023deep}, this work is based on the spatial traffic distribution in which the traffic demand in a certain area can be predicted and seen as fixed during a certain period, even though the users keep moving and have dynamic demand. The performance metric of satisfaction rate (SR), introduced by \cite{hatoum2013cluster}, is employed to evaluate the degree of satisfaction for sensing and communication demand. However, rather than treating the user/terminal as a point with specific coordinates, this grid-based model considers the traffic demand generated from a defined area encompassing a range of coordinates. To address the limitations of the point-to-point communication model within this innovative context, we introduce robust optimization tools to maximize the minimum SR and employ the cardinality constrained uncertainty set to control the robustness. To the  best of our knowledge, this is the first work that introduce the robust optimization tools to the grid-based traffic model. The problem is then reformulated as a mixed integer linear programming (MILP) via duality theory and we propose an iterative linear programming (LP) rounding algorithm to solve it in polynomial time. Numerical results show that RABS can improve the system performance by 28.61\% on average compared to fixed small cells. 

\vspace{-0.2cm}
\section{Application Scenario and System Model}
\label{systemmodel}

The grid-based model is a widely used model to characterize the spatial traffic distribution \cite{lee2014spatial}, and was first introduced to aerial networks in \cite{mozaffari2018beyond}. To employ the grid-based model in this letter, an urban geographical area is divided into multiple grids, assuming that the traffic demand generated from each grid remains unchanged and known within a certain time interval, e.g., half an hour or an hour. Our research focuses on a specific time period and aims to determine the deployment and resource allocation of RABS during this epoch. It is worth noting that the inherent flying function of RABS allows them to relocate to other lampposts in response to changes in traffic patterns in subsequent epochs. Besides, the RABS works with a ominidirectional antenna to transmit ISAC signals and receive the scattered echoes reflected by targets\cite{qin2023deep}.

\begin{figure}[!t]
\setlength{\abovecaptionskip}{0cm}
\centering 
\subfigure[OFDM waveform for ISAC, radar and communication (Inspired by Fig.1 in \cite{liu2019robust}). ]{\label{OFDM_ISAC} 
\includegraphics[width=0.53\linewidth]{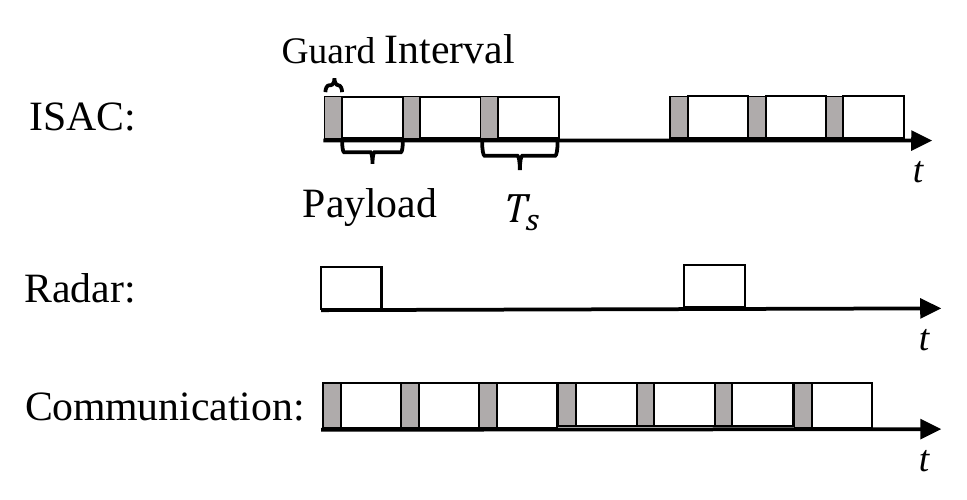}
} 
\subfigure[The longest and shortest distance between a grid and RABS.]{ 
\label{Bounds_distance}
\includegraphics[width=0.41\linewidth]{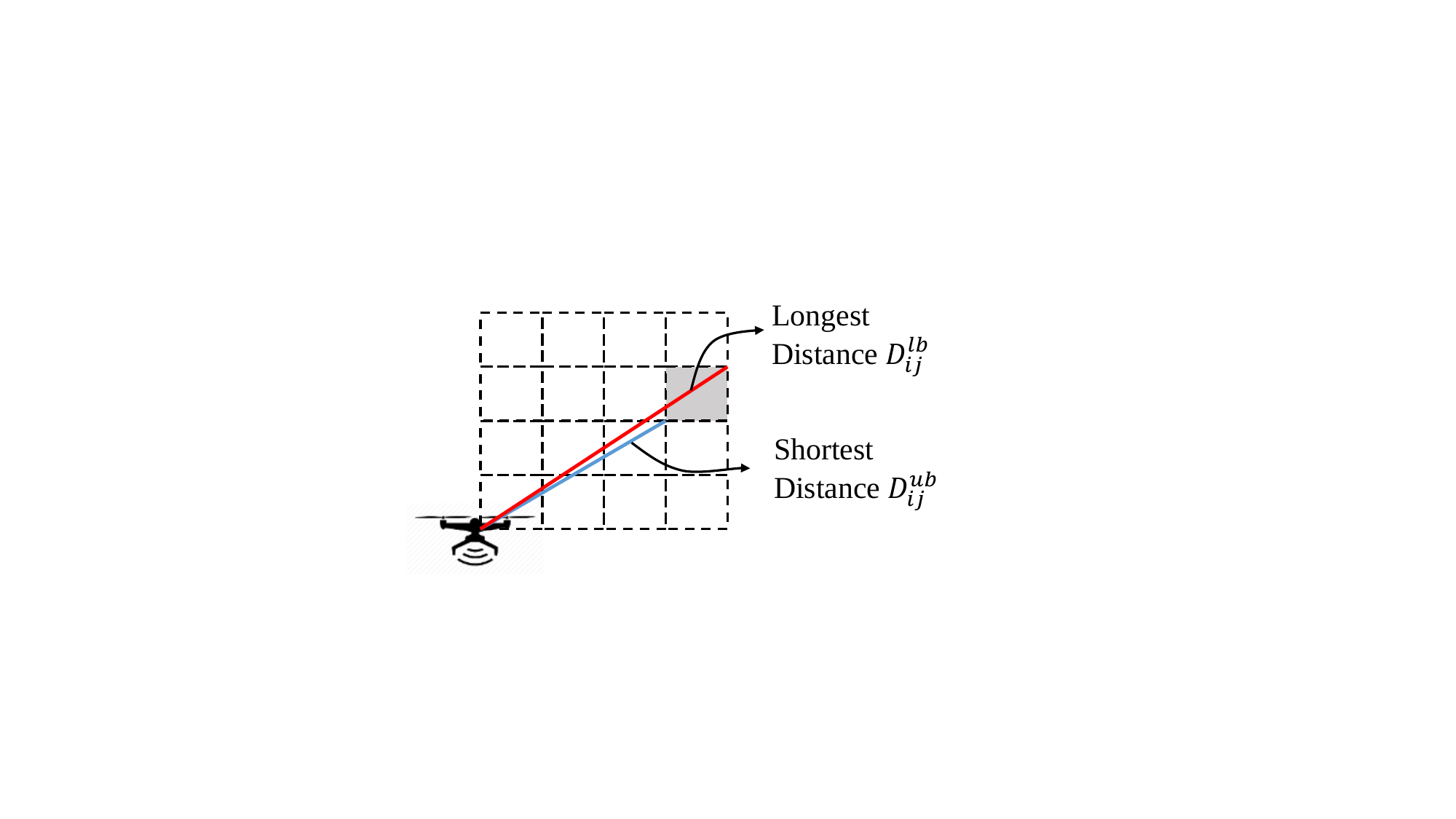}} 
\caption{Illustration of the system model.}
\vspace{-0.5cm}
\end{figure}

In addition to conventional communication functions, orthogonal frequency division multiplexing (OFDM) waveform is adopted for radar sensing applications because of its high spectrum efficiency, modulation flexibility and strong tolerance for inter-symbol interference. Unlike OFDM communication waveform which is continuous and consists of communication information and guard interval, OFDM sensing waveform is in the form of pulse signals without any embedded information or guard interval. To further improve the spectrum efficiency, OFDM-based ISAC applies a pulse OFDM waveform consisting of communication information to perform ISAC functions \cite{liu2019robust,liu2017adaptive,shi2018low}. The comparison of these three kinds of waveform is shown in Fig. \ref{OFDM_ISAC}. Specifically, suppose there are $K$ available OFDM subcarriers, denoted by $\mathcal{K} = \{1,2,...,K\}$, are utilized to perform ISAC. Therefore, the sensing signal performed on the subcarrier $k$ with $M$ consecutive integrated OFDM symbols can be described as \cite{liu2017adaptive,liu2019robust},
\begin{equation}
\label{Sen_signal}
\begin{aligned}
\! s_k(t) \! = \! e^{j2\pi f^c_k t} \! \sum_{n=0}^{N_s-1} \! \!  a_k c_{kn} e^{j2\pi k \Delta f (t - nT_s)} \! \cdot \! {\rm rect}[\frac{t \! - \! nT_s}{T_s}],
\end{aligned} 
\end{equation}
where $t$ is the continuous-time independent variable, $f^c_k$ and $\Delta f$ are the frequency and bandwidth of subcarrier $k$, $a_k$ and $c_{kn}$ denotes the amplitude and and phase code, respectively, $T_s$ is the duration of each completed OFDM symbol including both the guard intervals and elementary symbol, and ${\rm rect}[x]$ is the rectangle function that is equal to one when $x \! \in \! [0,1]$, and zero, otherwise. Accordingly, supposing the impulse response of a sensing target on subcarrier $k$, including path loss and radar cross section, is characterized by $h_k(t)$, the received signals can be written as $u_k(t) =  h_k(t)*s_k(t) + n(t)$. We consider a RABS that can be deployed in a certain area, which is divided into $I$ grids denoted by the set $\mathcal{I} = \{1,2,...,I\}$. There are a group of candidate locations distributed in that geographical area which can be chosen by RABSs for grasping; this set is denoted by $\mathcal{J} = \{1,2,...,J\}$. Besides, we should notice that one grid can be provisioned by one or multiple subcarriers while one subcarrier can only be assigned to at most one grid to avoid intra-cell interference.

Different performance metrics are employed to evaluate the sensing performance in aerial networks, such as Cramér–Rao lower bound and range resolution. In order to investigate the impact of RABS deployment and bandwidth allocation on the performance of ISAC systems, we utilize the conditional MI metric to assess the radar performance, similar to \cite{liu2017adaptive,liu2019robust,shi2018low,shi2019joint,nguyen2021hierarchical}. The conditional MI enables the characterization of the information-theoretic boundaries of the target information conveyed by the reflected sensing signal, which is commonly referred as sensing rate. Derived from \eqref{Sen_signal}, when the sensing demand generated from grid $i$ is served by a RABS deployed at candidate location $j$ and operating on the subcarrier $k$, the lower bound value of MI will be achieved if there is a user, distributed in grid $i$, having the worst channel gain \cite{liu2017adaptive},
\begin{equation}
\label{LB_MI_equation}
\begin{aligned}
M^{lb}_{ijk} = \frac{1}{2} \Delta f T_s N_s \log_2 ( 1 + |a_k|^2 T_s^2 N_s H^{sen,lb}_{ijk}/\sigma^2),
\end{aligned} 
\end{equation}
where $|a_k|^2$ calculates the transmission power of the subcarrier $k$, and $H^{sen,lb}_{ijk}$ represents the lower bound of the path loss value of the surveillance channel calculated by \cite{shi2018low}, 
\begin{equation}
\label{LB_Sen_channel}
\begin{aligned}
H^{sen,lb}_{ijk} = G_t^s G_r^s\eta \lambda_k^2/ \big((4 \pi)^3 {D^{lb}_{ij}}^4\big),
\end{aligned} 
\end{equation}
where $G_t^s$ and $G_r^s$ is the transmitting and receiving antenna gain, respectively, $\eta$ denotes the mean of radar cross-section of the targets distributed in the grid, $\lambda_k$ is the wavelength in the subcarrier $k$ that could be calculated by $\lambda_k = c/f^c_k$ where $c$ is the speed of light, $D^{lb}_{ij}$ denotes the longest distance between the grid $i$ and candidate location $j$. Similarly, introducing the shortest distance $D^{ub}_{ij}$ into \eqref{LB_MI_equation} and \eqref{LB_Sen_channel}, we can calculate the upper bound value of channel gain and MI in the best case, denoted by $H^{sen,ub}_{ijk}$ and $M^{ub}_{ijk}$. An illustration of the lower and upper bounds of the distance is shown in Fig. \ref{Bounds_distance}. For notational convenience, we calculate the average MI as $\overline{M}_{ijk}=(M^{ub}_{ijk}+M^{lb}_{ijk})/2$ and bias as $\hat{M}_{ijk}=(M^{ub}_{ijk}-M^{lb}_{ijk})/2$. Consequently, for any user distributed in the grid $i$, the MI should take value from the range $[\overline{M}_{ijk}-\hat{M}_{ijk}, \overline{M}_{ijk}+\hat{M}_{ijk}]$.

Moreover, the data rate is applied as the metric to evaluate the communication performance. The lower bound of the achievable rate can be calculated by,
\begin{equation}
\label{LB_Rate_equation}
\begin{aligned}
R^{lb}_{ijk} = \Delta f \log_2 ( 1 + |a_k|^2 H^{com,lb}_{ijk}/\sigma^2 ),
\end{aligned} 
\end{equation}
where $H^{com,lb}_{ijk}$ indicates the lower bound of the communication channel gain calculated as follows\cite{shi2018low,shi2019joint}, 
\begin{equation}
\label{LB_Com_channel}
\begin{aligned}
H^{com,lb}_{ijk} = G_t^c G_r^c \lambda_k^2/ \big((4 \pi)^2 {D^{lb}_{ij}}^2\big),
\end{aligned} 
\end{equation}
where $G_t^c$ and $G_r^c$ is the transmitting and receiving antenna gain.\footnote{Similar as \cite{shi2018low,shi2019joint}, we employ the free-space channel model for simplicity. Other models can be employed in the proposed formulation straightforwardly. } It is worth pointing out that the concept of the worst channel gain is investigated in \cite{zhang2022reliable} for reliable communications. The upper bound of communication channel gain and data rate, denoted by $H^{com,ub}_{ijk}$ and $R^{ub}_{ijk}$, can be then obtained by introducing the shortest distance $D^{ub}_{ij}$ into \eqref{LB_Rate_equation} and \eqref{LB_Com_channel}. The average rate and bias can be calculated as $\overline{R}_{ijk}=(R^{ub}_{ijk}+R^{lb}_{ijk})/2$ and $\hat{R}_{ijk}=(R^{ub}_{ijk}-R^{lb}_{ijk})/2$, respectively. Accordingly, the data rate for any user distributed in the grid $i$ would be within the range $[\overline{R}_{ijk}-\hat{R}_{ijk}, \overline{R}_{ijk}+\hat{R}_{ijk}]$.

\begin{linenomath}
\begin{figure*}
\begin{subequations}
\begin{empheq}[left={\empheqlbrace\,}]{align}
& \widetilde{M}_i \triangleq \frac{1}{M_i} \Big( \underbrace{\sum_{j \in \mathcal{J}} \sum_{k \in \mathcal{K}} \overline{M}_{ijk} x_{ijk}}_{\substack{\text{The total satisfied MI when} \\ \text{all grids have the average MI.}}}
- \!\!\!\!\! \underbrace{\max_{\{\mathcal{J}_i \times \mathcal{K}_i \cup (j_i,k_i) \,|\, \mathcal{J}_i \subseteq \mathcal{J}, \mathcal{K}_i \subseteq \mathcal{K}, \atop | \mathcal{J}_i \times \mathcal{K}_i| \leq \lfloor \Gamma_i \rfloor, (j_i,k_i) \in \mathcal{J} \times \mathcal{K} - \mathcal{J}_i \times \mathcal{K}_i  \}} \Big\{ \sum_{j \in \mathcal{J}_i} \sum_{k \in \mathcal{K}_i} \hat{M}_{ijk} x_{ijk}  + (\Gamma_i - \lfloor \Gamma_i \rfloor) \hat{M}_{ij_ik_i} x_{ij_ik_i} \Big\} \Big)}_{\substack{\text{Robust bias of the total satisfied MI which means up to $\lfloor \Gamma_i \rfloor$ of these coefficients are allowed to change to the} \\ \text{worst MI, and one coefficient changes by at most $(\Gamma_i - \lfloor \Gamma_i \rfloor) \hat{M}_{ij_ik_i}$. Defined as the protection function in \eqref{ProtectFunction}.}}},\label{sense_SR} \\
& \widetilde{R}_i \triangleq \frac{1}{R_i} \Big( \sum_{j \in \mathcal{J}} \sum_{k \in \mathcal{K}} \overline{R}_{ijk} y_{ijk} 
- \!\! \max_{\{\mathcal{J}_i \times \mathcal{K}_i \cup (j_i,k_i) \,|\, \mathcal{J}_i \subseteq \mathcal{J}, \mathcal{K}_i \subseteq \mathcal{K}, \atop | \mathcal{J}_i \times \mathcal{K}_i| \leq \lfloor \Lambda_i \rfloor, (j_i,k_i) \in \mathcal{J} \times \mathcal{K} - \mathcal{J}_i \times \mathcal{K}_i  \}} \Big\{ \sum_{j \in \mathcal{J}_i} \sum_{k \in \mathcal{K}_i} \hat{R}_{ijk} y_{ijk} + (\Lambda_i - \lfloor \Lambda_i \rfloor) \hat{R}_{ij_ik_i} y_{ij_ik_i} \Big\} \Big), \label{comm_SR}
\end{empheq}
\end{subequations}
\vspace{-0.65cm}
\end{figure*}
\end{linenomath}

Three sets of binary variables are utilized to formulate the subcarrier allocation, grid association and RABS deployment. Specifically, $x_{ijk} \in \{0,1\}$ indicates whether a RABS located at location $j$ performs sensing operations to the user $i$ by the subcarrier $k$ or not;  $y_{ijk} \in \{0,1\}$ denotes whether a RABS located at location $j$ communicate with the user $i$ on the subcarrier $k$ or not; $z_j \in \{0,1\}$ indicates whether the RABS would be deployed at location $j$. Because our objective is to satisfy these demands as much as possible under resource constraints, we employ the satisfaction rate (SR) to evaluate the degree of satisfaction for sensing and communication demand \cite{hatoum2013cluster}. Accordingly, the SR for sensing demand in grid $i$ is defined by \eqref{sense_SR} shown on the top of this page. A parameter $\Gamma_i$, normally called the protection level for the $i^{th}$ constraint, is introduced to control the conservatism of the robust optimization model. Specifically, in the numerator of \eqref{sense_SR}, the first part calculates the total served sensing MI when all grids perform the average channel gain, i.e., have the average MI. The second part is the robust bias, which indicates that there are up to $\lfloor \Gamma_i \rfloor$ coefficients allowed to change within the range $[\overline{M}_{ijk}-\hat{M}_{ijk}, \overline{M}_{ijk}+\hat{M}_{ijk}]$, and one coefficient can at most change by $(\Gamma_i - \lfloor \Gamma_i \rfloor) \hat{M}_{ijk}$. This kind of uncertainty set is referred as cardinality constrained uncertainty set in \cite{bertsimas2004price}, which reflects the inherent nature that only a subset of grids perform the worst channel gain in order to adversely affect the MI performance. Considering two extreme cases, setting $\Gamma_i = 0$ is the most ideal scenario when all grids have the average sensing performance. In contrast, setting $\Gamma_i = |\mathcal{J}_i \times \mathcal{K}_i|$ is the most conservative case in which all grids perform the worst channel gain and therefore have the lowest MI. Overall, the numerator in \eqref{sense_SR} calculates the satisfied sensing demand under the cardinality constrained uncertainty set and the denominator $M_i$ denotes the sensing demand of the grid $i$. Therefore, \eqref{sense_SR} defines the sensing SR $\widetilde{M}_i$. Similarly, the communication SR $\widetilde{R}_i$ is defined by \eqref{comm_SR} where $\Lambda_i$ and $R_i$ are the protection level and communication demand, respectively.

Hereafter, the proposed bi-objective optimization problem is formulated to maximize the weighted sum of minimum sensing and communication SR,
\begin{linenomath}
\begin{subequations}
\label{RO_formulation}
\begin{align}
\; & \max_{\mathbf{X} , \mathbf{Y}, \mathbf{Z},\widetilde{M},\widetilde{R}} \; \mu \widetilde{M} + (1-\mu) \widetilde{R} \label{Primal_obj} \\
s.t.
\; &  \widetilde{M}_i \geq \widetilde{M}, \; \widetilde{R}_i \geq \widetilde{R}, \quad \forall i, \label{min_SR}\\
\; &  \sum_{i \in \mathcal{I}} \sum_{j \in \mathcal{J}} x_{ijk} \leq 1, \; \sum_{i \in \mathcal{I}} \sum_{j \in \mathcal{J}} y_{ijk} \leq 1, \quad \forall k, \label{onecarrier} \\
\; &  \sum_{i \in \mathcal{I}}\sum_{k \in \mathcal{K}} x_{ijk} \leq IK z_j, \; \sum_{i \in \mathcal{I}}\sum_{k \in \mathcal{K}} y_{ijk} \leq IK z_j, \; \; \forall j, \label{onlyrabs} \\
\; &  \sum_{j \in \mathcal{J}} z_j \leq 1, \label{num_RABS} \\
\; &  x_{ijk}, y_{ijk}, z_{j} \in \{0,1\}, \quad \forall i,j,k, \\
\; & \widetilde{M}, \widetilde{R} \in [0,1], \label{nonnegative}
\end{align}
\end{subequations}
\end{linenomath}
where $\mathbf{X} \triangleq \{x_{ijk}\}$, $\mathbf{Y} \triangleq \{y_{ijk}\}$ and $\mathbf{Z} \triangleq \{z_{j}\}$ are the set of variables, $\mu \!\in\! [0,1]$ is a predefined weight parameter. Eq. \eqref{min_SR} denotes the minimum sensing and communication SR by $\widetilde{M}$ and $\widetilde{R}$. The constraints in \eqref{onecarrier} denote that each orthogonal subcarrier can at most allocated to one grid for sensing or communication to avoid intra-cell interference, respectively. Eq. \eqref{onlyrabs} ensures that only when a RABS has been deployed at the location $j$ then the grids can be associated to it for joint sensing and communication. Eq. \eqref{num_RABS} indicates that there is only one RABS that can be deployed. 

\vspace{-0.2cm}
\section{MILP Reformulation and Algorithm Design}
\label{reformulationalgorithm}

\subsubsection{MILP Reformulation}

To convert the constraints in \eqref{sense_SR} into linear constraints, we first define the protection function with a given $\mathbf{X}^*$ as,
\begin{equation}
\label{ProtectFunction}
\begin{aligned}
\gamma_i(\mathbf{X}^*) = & \!\!\!\!\!\!  \!\!\!\!\!\! \max_{\{\mathcal{J}_i \times \mathcal{K}_i \cup (j_i,k_i) \,|\, \mathcal{J}_i \subseteq \mathcal{J}, \mathcal{K}_i \subseteq \mathcal{K}, \atop | \mathcal{J}_i \times \mathcal{K}_i| \leq \lfloor \Gamma_i \rfloor, (j_i,k_i) \in \mathcal{J} \times \mathcal{K} - \mathcal{J}_i \times \mathcal{K}_i  \}} \!\!\! \Big\{ \sum_{j \in \mathcal{J}_i} \sum_{k \in \mathcal{K}_i} \hat{M}_{ijk} x^*_{ijk}  \\
& + (\Gamma_i - \lfloor \Gamma_i \rfloor) \hat{M}_{ij_ik_i} x^*_{ij_ik_i} \Big\} ,
\end{aligned} 
\end{equation}
which can be written as the following problem:
\begin{linenomath}
\begin{subequations}
\label{ProtectProblem}
\begin{align}
\gamma_i(\mathbf{X}^*) = \; & \max_{\mathbf{w}_i} \; \sum_{j \in \mathcal{J}} \sum_{k \in \mathcal{K}} \hat{M}_{ijk} x^*_{ijk} w_{ijk} \\
s.t.
\; &  \sum_{j \in \mathcal{J}} \sum_{k \in \mathcal{K}} w_{ijk} \leq \Gamma_i,  \\
\; &  0 \leq  w_{ijk} \leq 1, \; \forall j,k,
\end{align}
\end{subequations}
\end{linenomath}
where $\mathbf{w}_i \triangleq \{ w_{ijk} | \forall j \! \in \! \mathcal{J}, \forall k \! \in \! \mathcal{K}\}$ is the introduced variable. The equality between \eqref{ProtectFunction} and \eqref{ProtectProblem} can be proved by the observation that the optimal solution of \eqref{ProtectProblem} must include $\lfloor \Gamma_i \rfloor$ variables taking the value of one and one variable at $\Gamma_i - \lfloor \Gamma_i \rfloor$. The detailed proof can be found in the Proposition 1 of \cite{bertsimas2004price}. Write the dual of \eqref{ProtectProblem} as follows,
\begin{linenomath}
\begin{subequations}
\label{dual_ProtectProblem}
\begin{align}
\gamma_i(\mathbf{X}^*) = \!\!\!\!\!\! & \min_{\alpha_i, \{\beta_{ijk} \,|\, \forall j, k\}} \; \sum_{j \in \mathcal{J}} \sum_{k \in \mathcal{K}} \beta_{ijk} + \Gamma_i \alpha_i \\
s.t.
\; &  \alpha_i + \beta_{ijk} \geq \hat{M}_{ijk} x^*_{ijk},  \; \forall j,k, \\
\; &  \alpha_i \geq 0, \; \beta_{ijk} \geq 0, \; \forall j,k,  
\end{align}
\end{subequations}
\end{linenomath}
where $\{\alpha_i\}$ and $\{\beta_{ijk}\}$ are dual variables. It can be observed that the problem (8) is a linear programming thus the strong duality is held between \eqref{ProtectProblem} and \eqref{dual_ProtectProblem}, i.e., they have the equal optimal solutions if feasible. Introducing \eqref{dual_ProtectProblem} into \eqref{sense_SR}, the constraints $\widetilde{M}_i \geq \widetilde{M}$ in \eqref{min_SR} can be rewritten as the following constraint set:
\begin{linenomath}
\begin{subequations}
\label{minsense_reform}
\begin{empheq}[left={\empheqlbrace\,}]{align}
&  \frac{1}{M_i} \Big( \sum_{j \in \mathcal{J}} \sum_{k \in \mathcal{K}} (\overline{M}_{ijk} x_{ijk}
- \! \beta_{ijk}) \! - \!  \Gamma_i \alpha_i \Big)
\geq \widetilde{M}, \; \forall i,  \label{minsense_reform_head} \\
&  \alpha_i + \beta_{ijk} \geq \hat{M}_{ijk} x_{ijk}, \; \forall i,j,k, \\
&  \alpha_i \geq 0, \; \beta_{ijk} \geq 0, \; \forall i,j,k. \label{minsense_reform_end}
\end{empheq}
\end{subequations}
\end{linenomath}

Applying the same procedure to  the constraints $\widetilde{R}_i \geq \widetilde{R}$ in \eqref{min_SR}, the problem \eqref{RO_formulation} can be then reformulated as a MILP without loss of optimality.

\subsubsection{Iterative LP Rounding Algorithm}
\label{LP_rounding}
To overcome the curse of dimensionality, an iterative LP rounding algorithm proposed in \cite{chen2021efficient} is employed to solve the reformulated MILP problem approximately. Firstly, we focus on a selected location and use the same method to traverse all candidate locations at subsequent stages to obtain the best one. It can be observed that the sensing and communication decisions in \eqref{RO_formulation} can be decoupled once the variable $\mathbf{Z}$ is determined. We set $z_{j'} = 1$ and all other elements in $\mathbf{Z}$ are zero. A MILP problem including only the variables related to the sensing task can be written from \eqref{RO_formulation} and \eqref{minsense_reform} as,
\begin{linenomath}
\begin{subequations}
\label{Sensing_MILP}
\begin{align}
\; & \max_{\mathbf{X_{j'}}, \widetilde{M}, \mathbf{A}, \mathbf{B_{j'}}} \; \mu \widetilde{M} \label{Sensing_MILP_obj} \\
s.t.
\; &  \frac{1}{M_i}\Big( \sum_{k \in \mathcal{K}} (\overline{M}_{ij'k} x_{ij'k}
-  \beta_{ij'k}) \! - \! \Gamma_i \alpha_i \Big)
\geq \widetilde{M}, \, \forall i, \label{Sensing_MILP_con1} \\
\; & \alpha_i + \beta_{ij'k} \geq \hat{M}_{ij'k} x_{ij'k},  \; \forall i,k,  \label{Sensing_MILP_con2} \\
\; &  \sum_{i \in \mathcal{I}} x_{ij'k} \leq 1, \; \forall k, \label{Sensing_MILP_con3} \\
\; & \widetilde{M} \in [0,1], \;  \alpha_i \geq 0, \; \beta_{ij'k} \geq 0, \; \forall i,k, \label{Sensing_MILP_con4} \\
\; &  x_{ij'k}\in \{0,1\}, \; \forall i,k, \label{Sensing_MILP_con6} 
\end{align}
\end{subequations}
\end{linenomath}where $\mathbf{X_{j'}} \triangleq \{x_{ij'k} \big| \forall i \in \mathcal{I}, \forall k \in \mathcal{K}\}$, $\mathbf{A} \triangleq \{\alpha_{i}\}$ and $\mathbf{B} \triangleq \{\beta_{ij'k} \big| \forall i \in \mathcal{I}, \forall k \in \mathcal{K} \}$ are the sets of variables.

To apply the iterative LP rounding algorithm \cite{chen2021efficient}, we first solve the linear relaxation of the problem \eqref{Sensing_MILP}, that is, replacing the constraints in \eqref{Sensing_MILP_con6} by $x_{ij'k}\in [0,1]$, and denote the solution as $(\mathbf{X_{j'}^*}, \widetilde{M}^*, \mathbf{A}^*, \mathbf{B}^*)$. If $\mathbf{X_{j'}^*}$ is binary, the optimal solution for \eqref{Sensing_MILP} is obtained. Otherwise, we introduce $x_{ij'k} = 1$ to \eqref{Sensing_MILP} if $x_{ij'k}^* = 1$ as new constraints. Afterwards we would decide to round the variables with fractional values in $\mathbf{X_{j'}^*}$ to binary values via a procedure of verifying feasibility. Firstly, we select one variable with the largest fractional value in $\mathbf{X_{j'}^*}$ and denote it as $x_{i_0j'k_0}$.\footnote{Because the objective is to maximize the minimum SR, it is suggested that in this step, prioritize selecting the elements in $\mathbf{X_{j'}^*}$ corresponding to the grids that have not allocated any subcarriers to guarantee the fairness.} We add the constraint $x_{i_0j'k_0} = 1$ to \eqref{Sensing_MILP} and try to solve this modified LP. If it is infeasible, we set $x_{i_0j'k_0} = 0$ and round other variables according to $\mathbf{X_{j'}^*}$. If the modified LP is feasible, we add the constraint $x_{i_0j'k_0} = 1$ to \eqref{Sensing_MILP} and repeat the above procedure until a binary $\mathbf{X_{j'}}$ is achieved or there is no more subcarrier can be allocated. More details of the iterative LP rounding algorithm could be found in \cite{chen2021efficient}.

In the section 6.6.1 of \cite{ben2001lectures}, the worst case of solving a linear programming is $\mathcal{O}\big((n^v+n^c)^{1.5}{n^v}^2\big)$, where $n^v$ and $n^c$ are the number of variables and constraints, respectively. In the iterative LP rounding algorithm, the number of iterations is upper bounded by $I \times K$, thus the complexity of the proposed algorithm is approximately $\mathcal{O}\big(IK\cdot(n^v+n^c)^{1.5}{n^v}^2 \big)$, where $n^c = 2IK+I+K+1$ and $n^v$ is upper bounded by $2IK+I+1$ for the linear relaxation of \eqref{Sensing_MILP}.

\vspace{-0.2cm}
\section{Numerical Investigations}
\label{NumericalInvest}

A geographical area of $100 \times 100$ m$^2$  is divided into 25 small square grids with the size of $20 \times 20$ m$^2$, where 10 candidate locations distributed randomly for RABS grasping. The sensing and communication demand of grids follows the log-normal distribution \cite{wang2015approach}, where the mean value and standard deviation are denoted by $[m^{sen}, m^{com}]$ and $[\sigma^{sen}, \sigma^{com}]$ \cite{wang2015approach}. Hereafter, unless otherwise specified, we set $m^{sen} = 15$ bit, $m^{com} = 20$ Mbps and $\sigma^{sen} = \sigma^{com} = 1$ unless otherwise stated. Moreover, the carrier frequency of the ISAC signals is $f^c_0 = 3$ GHz and each subcarrier has the spacing $\Delta f = 0.25$ MHz. Accordingly, the frequency of the $k^{th}$ subcarrier is calculated by $f_k^c = f^c_0 + k\Delta f$ \cite{liu2017adaptive}. For notational convenience, we introduce a robustness parameter $\delta$ to control the protection level $\{\Gamma_i\}$ and $\{\Lambda_i\}$, that is, $\Gamma_i = \Lambda_i = \delta \times J \times K$. Taking $\delta = 10^{-1}$ as an example, it means that 10\% of the coefficients in \eqref{sense_SR}-\eqref{comm_SR} are allowed to take values from $[\overline{M}_{ijk}-\hat{M}_{ijk}, \overline{M}_{ijk}+\hat{M}_{ijk}]$ and $[\overline{R}_{ijk}-\hat{R}_{ijk}, \overline{R}_{ijk}+\hat{R}_{ijk}]$. Other simulation parameters are reported in Table \ref{TAB para}. 

\label{NumericalResults}
\begin{table}[!t]
\centering
\caption{Parameter Settings}
\label{TAB para}
\begin{tabular}{ll|ll}
\hline
Parameter & Value & Parameter & Value\\
\hline
$K$ & 64 & $T_s$ & 5 $\rm \mu$s  \\
$G_t^s,\,G_r^s$ & 30 dB \cite{shi2018low} & $N_s$ & 16 \cite{liu2019robust} \\
$G_r^c$ & 30 dB \cite{shi2018low} & $\sigma^2$ & -174 dBm/Hz  \\
$G_s^c$ & 0 dB & $|a_k|^2$ & 1 W \\
$\eta$ & 1 m$^2$ \cite{qin2023deep} &  $\mu$ & 0.5 \\
\hline
\end{tabular}
\vspace{-0.2cm}
\end{table}

\begin{figure}[!t]
\setlength{\abovecaptionskip}{-0.1cm}
\centering
\includegraphics[width=0.66\linewidth]{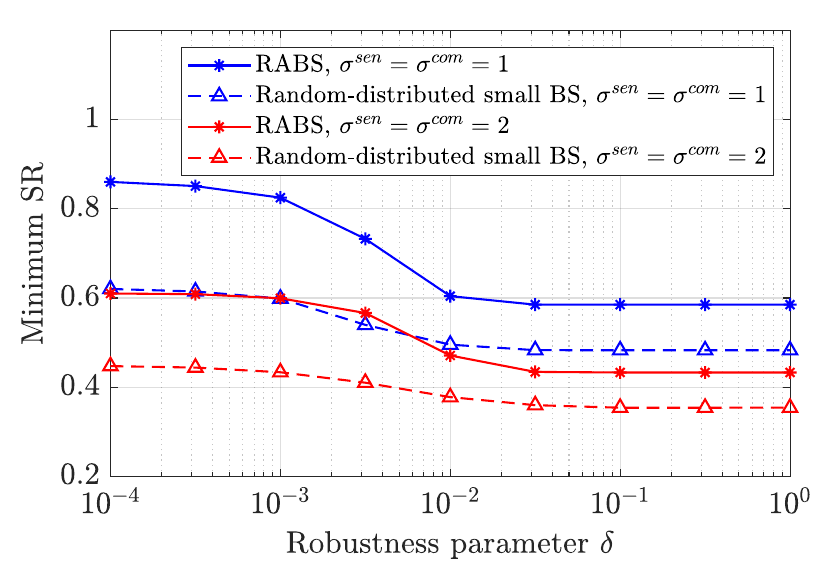}
\caption{Achievable minimum SR (objective function) versus robustness.}
\label{diff_robustness}
\vspace{-0.4cm}
\end{figure}

By adjusting the robustness parameter $\delta$, we control the protection level $\{\Gamma_i\}$ and $\{\Lambda_i\}$ as well as the robustness of the problem \eqref{RO_formulation}. It can be observe from Fig. \ref{diff_robustness} that the minimum SR decreases as the robustness increases. This is in accordance with the intuition that the growth of system robustness comes at the expense of system performance. Taking a fixed small cell distributed randomly as a benchmark, it is shown from Fig. \ref{diff_robustness} that the RABS can improve the system performance by 28.61\% and 21.46\% on average when setting the standard deviation to 1 and 2, respectively. Moreover, comparing the results for different standard deviation values of traffic distribution, it can be seen that the robustness has less impact on the system performance when the traffic spatial distribution is highly heterogeneous, represented as smoother curves in Fig. \ref{diff_robustness}.

\begin{figure}[!t]
\setlength{\abovecaptionskip}{-0.1cm}
\centering
\includegraphics[width=0.66\linewidth]{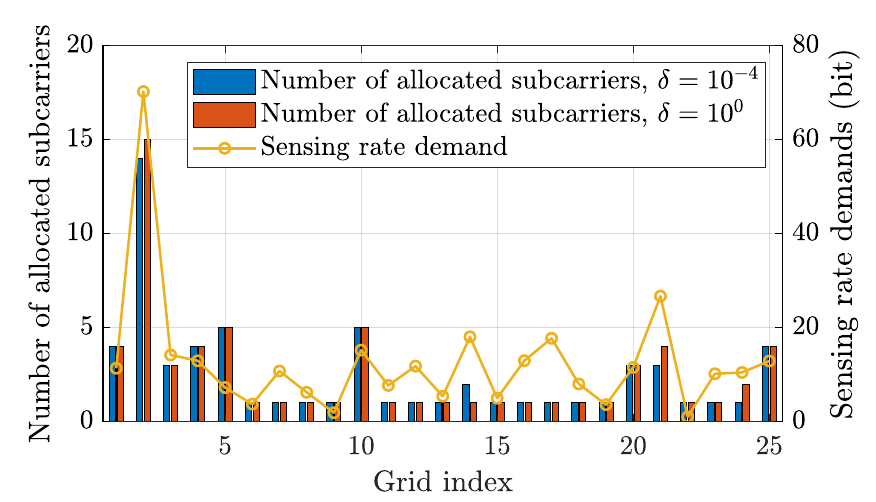}
\caption{Subcarrier allocation versus sensing rate demand.}
\label{SC_allocate}
\vspace{-0.4cm}
\end{figure}

Fig. \ref{SC_allocate} investigates the number of subcarriers allocation versus the sensing traffic distribution. Note that more subcarriers are biased to grids with higher traffic demand. The reason is that our objective is to maximize the minimum SR to guarantee fairness. Moreover, Fig. \ref{SC_allocate} shows that the robustness parameter $\delta$ also affects the subcarrier allocation decisions. Comparing the results when setting $\delta = 10^{-4}$ and $\delta = 10^0$, the number of allocated subcarriers differs in grids 2, 14, 21, and 24.

\begin{figure}[!t]
\setlength{\abovecaptionskip}{0cm}
\centering 
\subfigure[Convergence behaviour of the proposed algorithm. ]{\label{conv_alg} 
\includegraphics[width=0.67\linewidth]{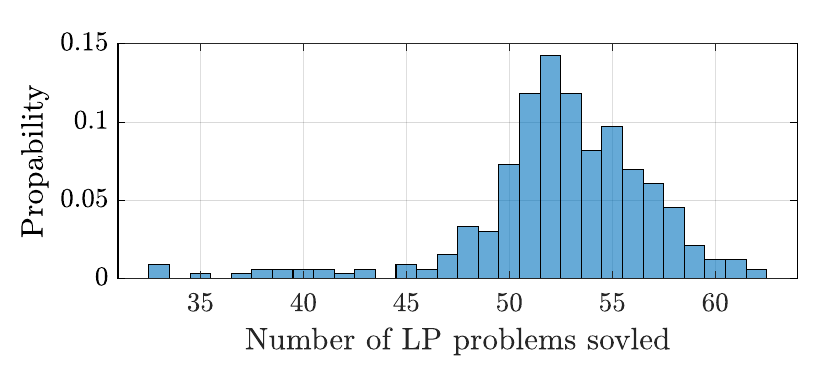}
} 
\subfigure[Optimality gap of the proposed algorithm.]{ 
\label{opt_alg}
\includegraphics[width=0.67\linewidth]{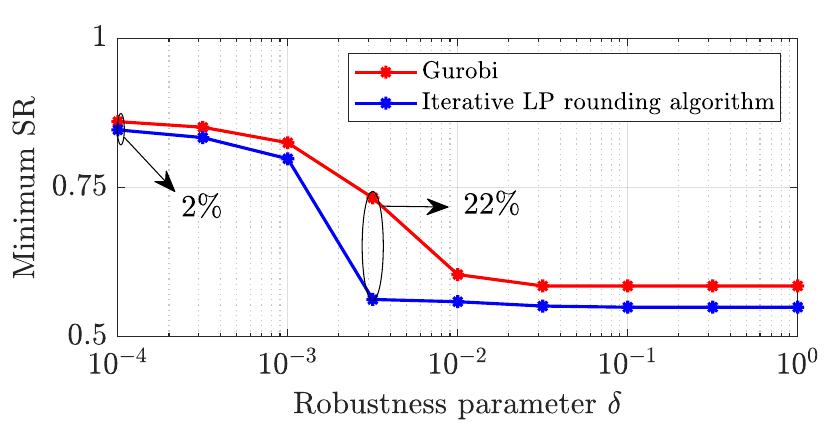}} 
\caption{Performance analysis of the iterative LP rounding algorithm.}
\label{per_alg} 
\vspace{-0.4cm}
\end{figure}

The performance of the proposed iterative LP rounding algorithm is analyzed in Fig. \ref{per_alg}. Although the maximum number of iterations is upper-bounded by $I \! \times \! K$, as alluded in section \ref{LP_rounding}, in reality the stopping criteria is satisfied after solving a limited number of LP problems as shown in Fig. \ref{conv_alg}. Moreover, Fig. \ref{opt_alg} presents the optimal gap of the iterative LP rounding algorithm by comparing with the globally optimal solution solved by Gurobi \cite{gurobi}. Numerically, the optimality gap of the proposed method is at least 2\% when the robustness parameter is $10^{-4}$, and 22\% at most when robustness parameter is $10^{-2.5}$. However, as shown in section \eqref{LP_rounding}, the complexity of the proposed algorithm is in polynomial time, in contrast to the exponential worst-case complexity of Gurobi \cite{gurobi}.

\vspace{-0.2cm}
\section{Conclusions}
\label{Conclusions}

In this paper, a flexible integrated sensing and communication (ISAC) system is proposed, assisted by the robotic aerial base station (RABS). To characterize the users' mobility and changing demand, we employ a grid-based model to represent the spatial traffic distribution. A robust programming is formulated on the cardinality constrained uncertainty set to determine the RABS deployment and resource allocation. which is reformulated as a MILP via duality theory and solved by a proposed iterative LP rounding algorithm in polynomial time. Numerical investigations show that the minimum SR can be improved by 28.61\% on average thanks to the flexible mobility of RABS deployment. Future extensions of this letter may consider employing the novel orthogonal time frequency space modulation to improve the performance of OFDM-based ISAC systems \cite{yuan2021integrated}, and applying the Cramér–Rao lower bound as the metric to investigate how the maneuverability of the RABS can enhance the ISAC performance.

\vspace{-0.4cm}
\bibliographystyle{IEEEtran}
\bibliography{IEEEabrv,reference}

\end{document}